\documentclass[11pt,a4paper]{article}

\usepackage{graphicx}
\usepackage{color}
\usepackage{amsmath,amssymb}
\usepackage{bbold}
\usepackage{t1enc}
\usepackage{cite}

\usepackage{amsmath,amssymb}
\usepackage{amsfonts}
\usepackage{bbold}
\usepackage{latexsym} 
\usepackage{cancel}
\usepackage{graphicx, rotating}
\usepackage{epsfig}
\usepackage{color}
\usepackage[dvipsnames]{xcolor}

\DeclareMathSymbol{\shortminus}{\mathbin}{AMSa}{"39}
\DeclareMathSymbol{\shm}{\mathbin}{AMSa}{"39}
\newcommand{\oh}{\frac{1}{2}}

\newcommand{\RE}{\operatorname{Re}}
\newcommand{\IM}{\operatorname{Re}}

\newcommand{\app}{a_{1 \oh}}
\newcommand{\azp}{a_{0 \oh}}
\newcommand{\azm}{a_{0 -\oh}}
\newcommand{\amm}{a_{-1-\oh}}
\newcommand{\aSp}{a_\frac{1}{2}}
\newcommand{\aSm}{a_{-\frac{1}{2}}}

\begin{document}

\begin{center}
\begin{Large}
{\bf Full quantum tomography of top quark decays}
\end{Large}

\vspace{0.5cm}
\renewcommand*{\thefootnote}{\fnsymbol{footnote}}
\setcounter{footnote}{0}
J.~A.~Aguilar-Saavedra \\[1mm]
\begin{small}
Instituto de F\'isica Te\'orica IFT-UAM/CSIC, c/Nicol\'as Cabrera 13--15, 28049 Madrid, Spain \\
\end{small}
\end{center}

\begin{abstract}
Quantum tomography in high-energy physics processes has usually been restricted to the spin degrees of freedom. We address the case of top quark decays $t \to W b$, in which the orbital angular momentum ($L$) and the spins of $W$ and $b$ are intertwined into a 54-dimensional $LWb$ density operator. The entanglement between $L$ and the $W$ or $b$ spin is large and could be determined for decays of single top quarks produced at the Large Hadron Collider with Run 2 data. With the foreseen statistical and systematic uncertainties, the significance is well above $5\sigma$ from the separability hypothesis for $L$-$W$ entanglement, and $3.2\sigma$ for $L$-$b$. These would be the first entanglement measurements between orbital and spin angular momenta in high-energy physics. Likewise, the genuine tripartite entanglement between $L$ and the two spins could be established with more than $5\sigma$. The method presented paves the way for similar measurements in other processes.
\end{abstract}

\section{Introduction}

Quantum tomography plays a crucial role in our understanding of particle physics processes. By reconstructing the density operator, quantum tomography provides a detailed snapshot of the quantum properties of a system, allowing us to gain insight into its behaviour. The full knowledge of the density operator enables the investigation of quantum entanglement patterns, and can also be used as a novel probe for new physics. For high-energy physics processes, previous literature has focused on the spin degrees of freedom of top quarks~\cite{Kane:1991bg}, $W$~\cite{Aguilar-Saavedra:2015yza} and $Z$~\cite{Aguilar-Saavedra:2017zkn} bosons, or combinations of them~\cite{Rahaman:2021fcz,Ashby-Pickering:2022umy,Bernal:2023jba}, which can be determined from multi-dimensional angular distributions of their decay products. Likewise, spin entanglement tests have been proposed for top quarks~\cite{Afik:2020onf,Fabbrichesi:2021npl,Severi:2021cnj,Afik:2022kwm,Aguilar-Saavedra:2022uye,Afik:2022dgh,Dong:2023xiw,Han:2023fci}, muons~\cite{Aguilar-Saavedra:2023lwb}, $\tau$ leptons~\cite{Altakach:2022ywa}, weak bosons~\cite{Barr:2021zcp,Aguilar-Saavedra:2022wam,Aguilar-Saavedra:2022mpg,Fabbri:2023ncz,Ashby-Pickering:2022umy,Fabbrichesi:2023cev,Morales:2023gow}, and particles of different spin~\cite{Aguilar-Saavedra:2023hss,Aguilar-Saavedra:2024fig,Aguilar-Saavedra:2024hwd}. Spin-spin entanglement has been experimentally probed with top quark pairs by the ATLAS~\cite{ATLAS:2023fsd} and CMS~\cite{CMS:2024hgo} Collaborations.

Including the orbital angular momentum (hereafter referred to as $L$) is a non-trivial step further because momenta measurement prevents full direct access to its density operator $\rho_L$. High-energy physics experiments rely on the measurement of particle momenta, which projects over {\em diagonal} states. Specifically, when momenta $P$ are measured the resulting state is $\mathcal{P} \rho_L \mathcal{P}$, with $\mathcal{P} = |P \rangle \langle P |$, and different operators $\rho_L$ can lead to the same projection $\mathcal{P} \rho_L \mathcal{P}$. For the decay of top quarks and other two-body decays, the differential decay width is $d\Gamma \propto \langle \Omega | \rho_L | \Omega \rangle$, with $\Omega = (\theta,\phi)$ the flight direction of one of the decay products. $d\Gamma$ involves products of two spherical harmonics $\Pi_{ll'}^{mm'}(\theta,\phi) \equiv Y_l^m(\theta,\phi) Y_{l'}^{m'}(\theta,\phi)^*$ evaluated on the same arguments. These products $\Pi_{ll'}^{mm'}(\theta,\phi)$ are {\em not} orthogonal functions, therefore, different operators $\rho_L$ may result in the same angular distribution. An extreme case is a scalar, e.g. the Higgs boson, where the decay distribution is isotropic despite $\rho_L$ contains non-zero entries $|l\,m \rangle \langle l'\,m' |$, up to $l,l' =2$ in the decay to vector bosons and $l,l' =1$ in the decay to fermions.

In this Letter we consider a case of interest, the decay of top quarks $t \to W b$, to address the determination of the 54-dimensional $LWb$ density operator that completely describes angular momentum in the top quark decay. We make a novel use of the complementarity between `canonical' decay amplitudes, calculated using a fixed quantisation axis for all particles, and helicity amplitudes, in which the spin of the decay products is quantised in their direction of motion. While the former are the key tool to write down the $LWb$ density operator, the latter are very convenient to extract from multi-dimensional angular decay distributions the independent parameters that govern the top decay. The $LW$ and $Lb$ density operators are also of interest: they exhibit a large entanglement between the orbital and spin angular momentum parts, which can be experimentally determined, with a caveat that will be discussed. Previous experimental measurements of entanglement between orbital and spin angular momenta have been limited to photon pairs, see for example Refs.~\cite{photon1,Fickler:2016xig}. The genuine tripartite entanglement between $L$ and the two spins is also large, and could be experimentally established.

\section{Top quark decay amplitudes}
 
In the top quark rest frame we take a fixed reference system $(x,y,z)$ and parameterise the three-momentum of the $W$ boson as
\begin{equation}
\vec p = q (\sin \theta \cos \phi,\sin \theta \sin \phi,\cos \theta) \,.
\end{equation}
Within the helicity amplitude framework of Jacob and Wick~\cite{Jacob:1959at}, angular momentum conservation implies that for the $t \to W b$ decay the helicity amplitudes must have the form
\begin{equation}
A^h_{M \lambda_1 \lambda_2} (\theta,\phi) = a_{\lambda_1 \lambda_2} D_{M \lambda}^{1/2\,*} (\phi,\theta,0) \,,
\label{ec:aJW}
\end{equation}
with $M$ the third spin component of $t$, $\lambda_1$ and $\lambda_2$ the helicities of $W$ and $b$, respectively, and $\lambda = \lambda_1-\lambda_2$.  $D^J_{m'm}$ are the Wigner functions and the `reduced amplitudes' $a_{\lambda_1 \lambda_2}$ are constants that do not depend on the angles, the only non-zero ones being $\app$, $\azp$, $\azm$, $\amm$. The combinations $\lambda_1 = 1$, $\lambda_2 = -1/2$ and $\lambda_1 = -1$, $\lambda_2 = 1/2$ are not allowed by angular momentum conservation, because the $L$ component in the direction of motion $\vec p$ vanishes. Performing a change of basis for the $b$ spinors and the $W$ polarisation vectors one can obtain the canonical amplitudes $A^c_{M s_1 s_2} (\theta,\phi)$, in which the $W$ and $b$ spins are quantised in the $\hat z$ axis, with respective eigenvalues $s_1$ and $s_2$. Expanding them in terms of spherical harmonics $Y_l^m$, and bearing in mind that $\langle \Omega | l m \rangle = Y_l^m(\Omega)$, the amplitudes $A_{M s_1 s_2;lm}$ for decay to $|s_1 s_2 l m\rangle$ eigenstates can be identified. The non-zero ones are
\begin{align}
& A_{\oh 1 \oh;1-1} = A_{-\oh -1 -\oh;11} = \sqrt{\frac{\pi }{6}} \mathcal{F}_1^- \,, \notag \\ 
& A_{\oh 1\oh ;2-1} = - A_{-\oh -1 -\oh;21} =  \sqrt{\frac{\pi }{30}} \mathcal{F}_2 \,,  \notag \\ 
& A_{\oh 0 \oh;00} = - A_{-\oh 0 -\oh;00} =  \sqrt{\frac{\pi}{9}} \mathcal{F}_0  \,, \notag \\ 
& A_{\oh 0 \oh;10} = A_{-\oh 0 -\oh;10} = \sqrt{\frac{\pi }{3}} \mathcal{F}_1^0 \,, \notag \\  
& A_{\oh 0 \oh;20} = - A_{-\oh 0 -\oh;20} = - \sqrt{\frac{2 \pi }{45}} \mathcal{F}_2 \,, \notag \\ 
& A_{\oh -1 \oh;11} = A_{-\oh 1 -\oh;1-1} =  \sqrt{\frac{\pi }{6}} \mathcal{F}_1^+ \,, \notag \\ 
& A_{\oh -1 \oh;21} = - A_{-\oh 1 -\oh;2-1} = \sqrt{\frac{\pi }{30}} \mathcal{F}_2 \,,  \notag \\  
& A_{\oh 1 -\oh;00} = - A_{-\oh -1 \oh;00} =  - \sqrt{\frac{2 \pi}{9} }  \mathcal{F}_0 \,,  \notag \\ 
& A_{\oh 1 -\oh;10} = A_{-\oh -1 \oh;10} =  \sqrt{\frac{\pi }{3}} \mathcal{F} \,,  \notag \\ 
& A_{\oh 1 -\oh;20} = - A_{-\oh -1 \oh;20} = - \sqrt{\frac{\pi }{45}} \mathcal{F}_2 \,, \notag \\ 
& A_{\oh 0 -\oh;11} = A_{-\oh 0 \oh;1-1} =  - \sqrt{\frac{\pi }{3}} \mathcal{F} \,,  \notag \\
& A_{\oh 0 -\oh;21} = - A_{-\oh 0 \oh;2-1} = \sqrt{\frac{\pi }{15}} \mathcal{F}_2  \,, \notag \\
& A_{\oh -1 -\oh;22} = - A_{-\oh 1 \oh;2-2} = - \sqrt{\frac{2\pi}{15}} \mathcal{F}_2  \,.
\label{ec:amp5}
 \end{align}
We have introduced the combinations
\begin{align}
& \mathcal{F} = \amm - \app \,, \quad \mathcal{F}_1^0 = \azm - \azp \,, \notag \\
& \mathcal{F}_0 = \sqrt{2} ( \app + \amm ) + \azp + \azm \,, \notag \\
& \mathcal{F}_1^\pm = \pm (\amm - \app)  + \sqrt{2} ( \azp-  \azm ) \,, \notag \\
& \mathcal{F}_2 = \amm + \app -\sqrt{2} ( \azp + \azm)  \,.
\end{align}
We note that the dependence on angles $(\theta,\phi)$ is no longer present in the amplitudes (\ref{ec:amp5}) for decay to $|s_1 s_2 l m\rangle$ eigenstates. These amplitudes are completely general, and the relation with $a_{\lambda_1 \lambda_2}$ holds for any two-body decay of a spin-$1/2$ particle into massive spin-1 and spin-$1/2$ particles. For the top quark decay one can use the helicity amplitudes calculated in~\cite{Aguilar-Saavedra:2024fig} for a generic interaction
\begin{equation}
\mathcal{L} = - \frac{1}{\sqrt 2} \bar b \gamma^\mu (g_L P_L + g_R P_R) t \, W_\mu^- 
\label{ec:lagr}
\end{equation}
and match them to the general form (\ref{ec:aJW}), to obtain for this specific case the reduced amplitudes (or just amplitudes, for brevity)
\begin{align}
& \app = \frac{1}{\sqrt 2} (g_L H_1^- - g_R H_1^+ ) \,, \notag \\ 
& \azp = \frac{1}{\sqrt 2} (g_L H_1^- H_2^- - g_R H_1^+ H_2^+ ) \,, \notag \\
& \azm = \frac{1}{\sqrt 2} (g_L H_1^+ H_2^+ - g_R H_1^- H_2^- ) \,, \notag \\
& \amm = \frac{1}{\sqrt 2} (g_L H_1^+ - g_R H_1^- ) \,,
\label{ec:allpred}
\end{align}
with 
\begin{align}
& H_1^\pm = [m_t (E_b+m_b)]^{1/2} \left( 1 \pm \frac{q}{E_b+m_b} \right) \,, \notag \\
& H_2^\pm = \frac{E_W \pm q}{\sqrt{2} M_W} \,.
\end{align}
The masses of $t$, $W$ and $b$ are denoted as $m_t$, $M_W$ and $m_b$, as usual, and $E_W$, $E_b$ are the energies of $W$ and $b$, respectively, in the top quark rest frame. As a cross check, plugging (\ref{ec:allpred}) into (\ref{ec:amp5}) and using $\langle \Omega | l m \rangle = Y_l^m(\Omega)$, one recovers the canonical amplitudes $A_{M s_1 s_2} (\theta,\phi)$ obtained in~\cite{Aguilar-Saavedra:2024hwd} by direct calculation. 
Within the Standard Model (SM), and using $m_t = 172.5$ GeV, $M_W = 80.4$ GeV, $m_b = 4.8$ GeV,
one finds $(\app,\azp ,\azm,\amm) = g (5.4,1.8,231,153)$ GeV, with $g$ the electroweak constant. The amplitudes with $\lambda_2 = 1/2$ are suppressed because the $tbW$ interaction is left-handed and the $b$ quark is much lighter than the top quark.

\section{Density operators  for top quark decay}

Given a density operator parameterised as
\begin{equation}
 \rho_t = \frac{1}{2} \left( \! \begin{array}{cc} 1 + P_3 & P_1 - i P_2 \\ P_1 + i P_2 & 1 - P_3 \end{array} \! \right)
 \label{ec:rhot}
\end{equation}
describing the spin state of the decaying top quarks, the $LWb$ density operator $\rho_{LWb}$ for its decay products is 
\begin{equation}
(\rho_{LWb})_{s_1 s_2 ; l m}^{s_1' s_2' ; l' m'} = (\rho_t)_{MM'} A_{M s_1 s_2;lm} A_{M' s_1' s_2' ; l' m'}^* \,,
\label{ec:rhoLWb}
\end{equation}
with a sum over $M,M'$. Taking the partial trace over the $b$ spin, $W$ spin or $L$ subspaces we obtain the density operators $\rho_{LW}$, $\rho_{Lb}$ and $\rho_{Wb}$, respectively, with matrix elements
\begin{align}
& (\rho_{LW})_{s_1 ; l m}^{s_1' ; l' m'} = (\rho_t)_{MM'} A_{M s_1 s_2;lm} A_{M' s_1' s_2 ; l' m'} \,, \notag \\
& (\rho_{Lb})_{s_2 ; l m}^{s_2' ; l' m'} = (\rho_t)_{MM'} A_{M s_1 s_2;lm} A_{M' s_1 s_2' ; l' m'}^* \,, \notag \\
& (\rho_{Wb}) _{s_1 s_2}^{s_1' s_2'} = (\rho_t)_{MM'} A_{M s_1 s_2;lm} A_{M' s_1' s_2' ; l m}^* \,,
\end{align}
with implicit sums over repeated indices.
The operators $\rho_{LW}$ and $\rho_{Lb}$ have an obvious interpretation: they correspond to marginalising the $b$ quark or $W$ boson spin degrees of freedom. The interpretation of $\rho_{Wb}$ is less straightforward. Since in $\rho_{LWb}$ there is no angular dependence and the $L$ degrees of freedom are marginalised, this operator describes the $Wb$ spin degrees of freedom integrated over all top quark decay phase space. An independent computation of $\rho_{Wb}$ using the canonical decay amplitudes $A^c_{M s_1 s_2}$, followed by integration over $(\theta,\phi)$, renders exactly the same result for $\rho_{Wb}$.

For bipartite systems the entanglement can be characterised by
the Peres-Horodecki~\cite{Peres:1996dw,Horodecki:1997vt} criterion: for a density operator $\rho$ that describes a bipartite system $AB$, the positivity of the partial transpose over any subsystem, say $\rho^{T_B}$, is a necessary condition for separability. Conversely, a non-positive $\rho^{T_B}$ is a {\em sufficient} condition for entanglement. The amount of entanglement can be quantified by the negativity of the partial transpose on the $B$ subspace $\rho^{T_B}$~\cite{Plenio:2007zz},
\begin{equation}
N(\rho) = \frac{\| \rho^{T_B} \| - 1}{2} \,,
\label{ec:Nrho}
\end{equation}
where $\|X\| = \operatorname{tr} \sqrt{XX^\dagger} = \sum_i \sqrt{\lambda_i}$, where $\lambda_i$ are the (positive) eigenvalues of the matrix $XX^\dagger$. Because $\rho^{T_B}$ has unit trace, $N(\rho)$ equals the sum of the negative eigenvalues of $\rho^{T_B}$. (The result is the same when taking the partial transpose on subsystem $A$.) In the separable case $N(\rho) = 0$. As an example, in the SM we have for the decay of top quarks in a $S_3=1/2$ eigenstate $N(\rho_{LW}) = 0.615$, $N(\rho_{Lb}) = 0.396$, $N(\rho_{Wb}) = 0.010$, i.e. all the pairs of subsystems are entangled when tracing on the remaining one. Note, however, that $\rho_{Wb}$ is separable in the $m_b = 0$ limit, because in that case the $b$ quarks have left-handed helicity and there are only two non-zero $a_{\lambda_1 \lambda_2}$, those with $\lambda_2 = -1/2$. For unpolarised top quarks we have $N(\rho_{LW}) = 0.432$, $N(\rho_{Lb}) = 0.353$, $N(\rho_{Wb}) = 0$. The $W$-$b$ entanglement, which is zero or nearly zero in all situations, is ignored from now on.

The genuine tripartite entanglement can also be established. Namely, the fact that the state is not separable under any bipartition of the Hilbert space $\mathcal{H}_L \otimes \mathcal{H}_W \otimes \mathcal{H}_b$ in which $\rho_{LWb}$ acts. (For easier notation we label the Hilbert spaces for the $W$ and $b$ spins as $\mathcal{H}_W$ and $\mathcal{H}_b$, respectively.) For example, the $L$-$(Wb)$ entanglement can be tested by using the Peres-Horodecki sufficient condition with $A = \mathcal{H}_L$, $B = \mathcal{H}_W \otimes \mathcal{H}_b$, and the amount of entanglement can be quantified by the negativity $N$ in (\ref{ec:Nrho}). Likewise, the $W$-$(Lb)$ and $b$-$(LW)$ entanglement can be established. For the decay of top quarks in a $S_3 = 1/2$ eigenstate we have $N(\rho_{L[Wb]}) = 1.65$, $N(\rho_{W[Lb]}) = 0.80$, $N(\rho_{b[LW]}) =0.50$ for $L$-$(Wb)$, $W$-$(Lb)$ and $b$-$(LW)$ entanglement, respectively. For unpolarised top quarks we have $N = 1.24,0,59,0.49$.

\section{Parameter determination}

 Single top quarks are produced at the Large Hadron Collider (LHC) in the process $pp \to t \bar b j$ with a large polarisation in the direction of the spectator jet $j$. As shown in Ref.~\cite{Aguilar-Saavedra:2017wpl}, this allows to determine $a_{\lambda_1 \lambda_2}$, except for (i) a global normalisation that does not affect the density operators, and is set to unity; (ii) the relative phase $e^{i \alpha}$ between $\lambda_2 = 1/2$ and $\lambda_2 = -1/2$ amplitudes, which has a marginal effect on the entanglement because the former are quite suppressed. In addition to the $(\theta,\phi)$ angles previously introduced for the top quark decay, to measure $W$ spin observables one has to consider the angles $(\theta^*,\phi^*)$ describing the orientation of the charged lepton momentum in the $W$ rest frame, with respect to a $(x',y',z')$ system where the $\hat z'$ axis is in the direction of $\vec p$ and the $\hat y'$ angle lies in the $xy$ plane making an angle $\phi$ with the $\hat y$ axis.

The top quark polarisation, namely $P_{1-3}$ in Eq.~(\ref{ec:rhot}), is determined by using the charged lepton as spin analyser~\cite{Kane:1991bg}. We use forward-backward (FB) asymmetries on $\cos \theta_{\hat n}$, with $\theta_{\hat n}$ the angle between the charged lepton momentum in the top quark rest frame and an arbitrary axis $\hat n$. Their relation to the top quark polarisation $P_{\hat n}$ in that axis is
\begin{equation}
A_\text{FB}^{\hat n} = \frac{P_{\hat n}}{2} \,.
\end{equation}
For the determination of $a_{\lambda_1 \lambda_2}$ we use a simple set of observables. The well-known $W$ helicity fractions~\cite{Kane:1991bg} $F_{\pm}$, $F_0$ can be determined from a fit to the $\cos \theta^*$ distribution,
\begin{equation}
\frac{1}{\Gamma}\frac{d\Gamma}{d\cos \theta^*} = \frac{3}{8} F_+ (1+\cos\theta^*)^2 + \frac{3}{4} F_0 \sin^2 \theta^* 
+ \frac{3}{8} F_- (1-\cos\theta^*)^2 \,.
\end{equation}
Their relation to $a_{\lambda_1 \lambda_2}$ is
\begin{equation}
F_+ = |\app|^2 \,, \quad F_- = |\amm|^2 \,, \quad F_0 = |\azp|^2 + |\azm|^2 \,.
\end{equation}
The two amplitudes with $\lambda_1 = 0$ can be disentangled with a forward-backward central-edge asymmetry~\cite{Aguilar-Saavedra:2017wpl} on the quantity $\cos \theta (|\cos \theta^*| -t)$, with $t = (1 + \sqrt 2)^{1/3} - (1 + \sqrt 2)^{-1/3}$, 
\begin{equation}
A_\text{FB,EC}^{z,z'} = \frac{3}{2} (2t-1) P_3 \left[ |\azp|^2 - |\azm|^2 \right] \,.
\label{ec:afbec}
\end{equation}
The two phases between like-$\lambda_2$ amplitudes can be measured from FB asymmetries $A^{x'}$, $A^{y'}$, $A^{x'z'}$, $A^{y'z'}$ built on the quantities $\cos \phi^*$, $\sin \phi^*$, $\cos \phi^* \cos \theta^*$ and $\sin \phi^* \cos \theta^*$, respectively~\cite{Aguilar-Saavedra:2017wpl} 
\begin{align}
& A_\text{FB}^{x'} = -\frac{3\pi}{8 \sqrt 2} P_3 \RE \left[ \azp \app^* + \amm \azm^* \right] \,, \notag \\
& A_\text{FB}^{y'} = -\frac{3\pi}{8 \sqrt 2} P_3 \IM \left[ \azp \app^* + \amm \azm^* \right] \,, \notag \\
& A_\text{FB}^{x'z'} = -\frac{3\pi}{8 \sqrt 2} P_3 \RE \left[ \azp \app^* - \amm \azm^* \right] \,, \notag \\
& A_\text{FB}^{y'z'} = -\frac{3\pi}{8 \sqrt 2} P_3 \IM \left[ \azp \app^* - \amm \azm^* \right] \,.
\label{ec:afbxyz}
\end{align}
Finally, the relative phase between $\lambda_2 = 1/2$ and $\lambda_2 = -1/2$ amplitudes cannot be measured as long as the $b$ quark polarisation is not currently measured for top quark decays in the ATLAS and CMS experiments. However, its impact on entanglement is marginal because $\lambda_2 = 1/2$ amplitudes are quite suppressed.\footnote{In scenarios beyond the SM where $\lambda_2 = 1/2$ amplitudes are not supressed (e.g. in the presence of $\bar b_R \sigma^{\mu \nu} t_L$ or $\bar b_R \gamma^\mu t_R$ interactions) this would not be the case. The presence of such interactions could easily be detected by a non-vanishing value of $F_+$, and would also induce deviations in the asymmetries in (\ref{ec:afbec}) and (\ref{ec:afbxyz}) and polarisations $P_1$, $P_2$ from their SM prediction~\cite{Aguilar-Saavedra:2017wpl}.}
For top quarks in a $S_3 = 1/2$ eigenstate, and using the SM values for $a_{\lambda_1 \lambda_2}$ but inserting an arbitrary relative phase $e^{i \alpha}$ in $\app$ and $\azp$, $N(\rho_{LW})$ lies in the range $[0.614,0.628]$ and $N(\rho_{Lb})$ in the range $[0.396,0.411]$. The minimum entanglement is found for the SM value of the phase, $\alpha = 0$. This variation, of the order of the expected statistical uncertainty, is not relevant for the determination of the $L$-$W$ and $L$-$b$ entanglement. Similarly, the variation in the entanglement between different bipartitions of $\mathcal{H}_L \otimes \mathcal{H}_W \otimes \mathcal{H}_b$, for different values of $\alpha$, is unimportant. The maximum difference with respect to the SM is obtained for $\alpha = \pi$: $+0.025$ for $N(\rho_{L[Wb]})$ and $-10^{-3}$ for $N(\rho_{b[LW]})$, with $O(10^{-6})$ differences for $N(\rho_{W[Lb]})$.

\section{Experimental sensitivity}

We estimate the statistical sensitivity for the determination of the entanglement measures for bipartite and tripartite entanglement with the data collected at the LHC Run 2, working at the partonic level. An event sample is generated with {\scshape MadGraph}~\cite{Alwall:2014hca}, in the process $p p \to t \bar b j$, using NNPDF 3.1~\cite{NNPDF:2017mvq} parton density functions and setting as factorisation and renormalisation scale the average transverse mass. The resulting top quark polarisation is $P_3 = 0.845$ in the spectator jet axis, with $P_1 = 0.035$, $P_2 = 0$ in orthogonal directions~\cite{Aguilar-Saavedra:2014eqa,ATLAS:2022vym}. We perform pseudo-experiments, in which (i) a subset of events is randomly chosen; (ii) the values of $P_i$ and $a_{\lambda_1 \lambda_2}$ are determined from the set; (iii) the density operators are reconstructed using Eqs.~(\ref{ec:amp5}) and (\ref{ec:rhoLWb}); (iv) the entanglement measures are computed. For each pseudo-experiment we take a sample of 50000 events, which is the expected number of the events after selection cuts in the single top polarisation measurement performed by the ATLAS Collaboration~\cite{ATLAS:2022vym} with the full Run 2 luminosity (139 fb$^{-1}$).

There are two points that are worth being noted, concerning our procedure. First, the background from other processes is about the same size as the signal~\cite{ATLAS:2022vym}, and its statistical fluctuations will increase the signal uncertainty. However, since the stastistical uncertainties obtained with pseudo-experiments are well beyond $5\sigma$ sensitivity, this effect turns out to be irrelevant in order to establish the observation of entanglement. The second point concerns the relative phase that cannot be measured, which is set to its SM value, $\alpha = 0$. Using a non-SM value increases $N(\rho_{LW})$ up to $0.013$ and $N(\rho_{Lb})$ up to $0.017$. The impact of this phase on the entanglement between the different bipartitions of $\mathcal{H}_L \otimes \mathcal{H}_W \otimes \mathcal{H}_b$ is also negligible: $N(\rho_{L[Wb]})$ is increased by up to $0.030$ and $N(\rho_{b[LW]})$ decreases up to $-1.4 \times 10^{-3}$. Again, the effect is irrelevant in order to establish the observation of entanglement.

\begin{figure}[htb]
\begin{center}
\begin{tabular}{cc}
\includegraphics[height=5.4cm,clip=]{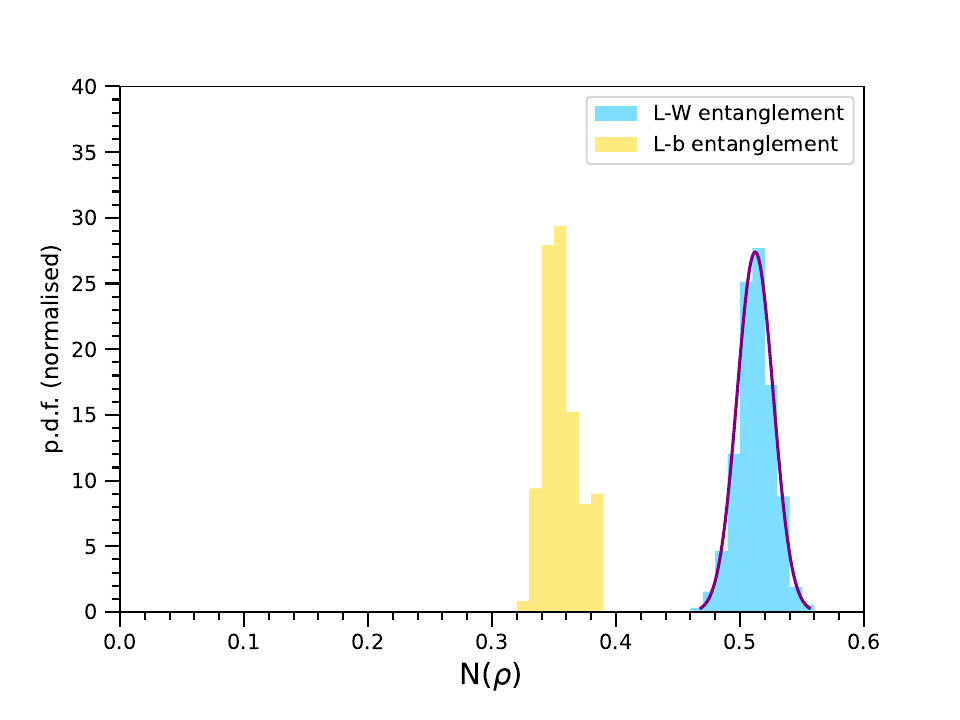} &
\includegraphics[height=5.4cm,clip=]{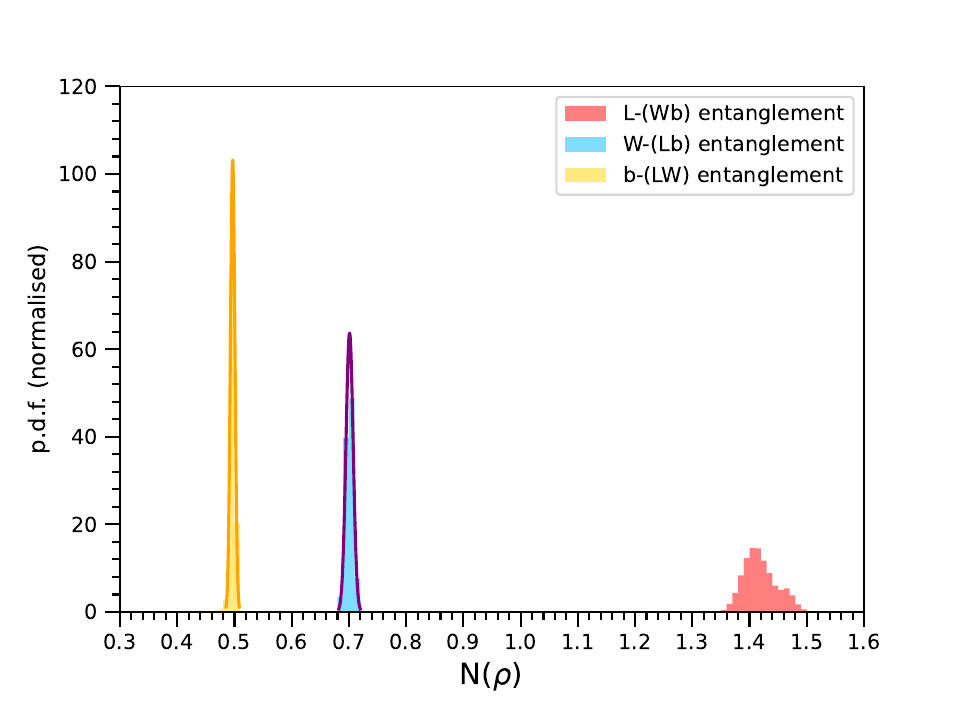}
\end{tabular} 
\caption{Distribution of entanglement measures $N$ obtained from pseudo-experiments, for bipartite entaglement (left) and tripartite entanglement (right). The curves show, in the interval $[-3\sigma,3\sigma]$, a Gaussian fit to the corresponding distribution.}
\label{fig:PE}
\end{center}
\end{figure}

For bipartite entanglement, the result of 20000 pseudo-experiments is presented in the left panel of Fig.~\ref{fig:PE}. For $L$-$W$ entanglement the distribution is Gaussian to a good approximation, and the expected statistical precision is $N(\rho_{LW}) = 0.512 \pm 0.015$. For $L$-$b$ entanglement the distribution presents a small shoulder caused by the boundary condition $F_+ \geq 0$. In this case, the mean is $\langle N(\rho_{Lb}) \rangle = 0.356$ and the standard deviation is $0.014$. In both cases, the measurements have a statistical precision that would allow to establish the entanglement ($N > 0$) beyond the $5\sigma$ criterion, nominally 35 and 25 standard deviations.

For tripartite entanglement, the result of 40000 pseudo-experiments is presented in the right panel of Fig.~\ref{fig:PE}. For $W$-$(Lb)$ and $b$-$(LW)$ the distributions are Gaussian, with $N(\rho_{W[Lb]}) = 0.701 \pm 0.006$, $N(\rho_{b[LW]}) = 0.496 \pm 0.004$. For $L$-$(Wb)$ the distribution has a small shoulder on the right. The mean and standard deviation are in this case $\langle N(\rho_{L[Wb]}) \rangle = 1.42$, $\sigma = 0.03$. The statistical difference from the separability hypothesis $N = 0$ is, in all cases, much larger than 5 standard deviations, and would allow to establish genuine tripartite entanglement.

The excellent statistical sensitivity expected for these entanglement measurements suggests that they will be limited by systematic uncertainties. For the single top polarisation, the Run 2 ATLAS measurement~\cite{ATLAS:2022vym}, which used a multi-dimensional template fit, has uncertainties of 17\%, 2.5\% and 10\%, for $P_1$, $P_2$ and $P_3$, respectively. The measurements of the $W$ helicity fractions by the CMS Collaboration in single top production with 20 fb$^{-1}$ of Run 1 data~\cite{CMS:2014uod} have systematic uncertainties below 4\%. Asymmetries equivalent to $A_\text{FB}^{x'}$, $A_\text{FB}^{y'}$, $A_\text{FB}^{x'z'}$, $A_\text{FB}^{y'z'}$ have been measured by the ATLAS Collaboration in single top production with 20 fb$^{-1}$ of Run 1 data~\cite{ATLAS:2017ygi}, finding systematics of 8\%, 3\%, 11\% and 5\%, respectively. 
A measurement of $A_\text{FB,EC}^{z,z'}$ has not been performed yet. It requires reconstruction of the $\hat z$ axis (spectator jet) and $\hat z'$ axis ($W$ momentum). In both cases the directions are well determined, so we expect a systematic uncertainty at the 10\% level, as obtained for $P_3$. Taking as central values the SM predictions and using the systematic uncertainties in Refs.~\cite{ATLAS:2022vym,CMS:2014uod,ATLAS:2017ygi} as benchmark, one can calculate the variation in entanglement observables due to (uncorrelated) variations in the input observables. The results are presented in Table~\ref{tab:syst}. Summing in quadrature all the different contributions, one can thus estimate the total systematic uncertainty on each entanglement measure, summarised in the last row of Table~\ref{tab:syst}. (This procedure does not take into account correlations between observables, which must be calculated once a simultaneous measurement of them all is performed.) 

As it can be drawn from Table~\ref{tab:syst}, the systematic uncertainty for $A_\text{FB,EC}^{z,z'}$, which is the observable for which a measurement is not yet available, only has a significant impact in $L$-$b$ entanglement. In this regard, it must be remarked that our assumption on its uncertainty, $A_\text{FB,EC}^{z,z'} = -0.17 \pm 0.1\;\text{(syst)}$, is rather conservative, and a better precision is likely. Summarising, our results for the statistical and systematic uncertainty of the several entanglement measures are
\begin{eqnarray}
N(\rho_{LW}) & = & 0.512 \pm 0.015\;\text{(stat)}~^{+0.054}_{-0.057}\;\text{(syst)} \,, \notag \\
N(\rho_{Lb}) & = & 0.356 \pm 0.014\;\text{(stat)}~^{+0.008}_{-0.11}\;\text{(syst)} \,, \notag \\
N(\rho_{L[Wb]}) & = & 1.42 \pm 0.03\;\text{(stat)}~^{+0.13}_{-0.09}\;\text{(syst)} \,, \notag \\
N(\rho_{W[Lb]}) & = & 0.701 \pm 0.006\;\text{(stat)}~^{+0.095}_{-0.038}\;\text{(syst)} \,, \notag \\
N(\rho_{b[LW]}) & = & 0.496 \pm 0.004\;\text{(stat)}~^{+0.026}_{-0.041}\;\text{(syst)} \,.
\end{eqnarray}
For all entanglement measures except $N(\rho_{Lb})$, the sensitivity to establish entanglement ($N > 0$) remains above the $5\sigma$ level after including the estimation of systematic uncertainties.

\begin{table}[t]
\begin{center}
\begin{tabular}{cccccc}
           & $N(\rho_{LW})$          & $N(\rho_{Lb})$            & $N(\rho_{L[Wb]})$      & $N(\rho_{W[Lb]})$      & $N(\rho_{b[LW]})$ \\
$P_1$ & $^{+0.01}_{-0}$          & $^{+0.003}_{-0}$        & $^{+0.03}_{-0}$           & $^{+0.044}_{-0.003}$ & $^{+0.021}_{-0.003}$ 
\\[1mm]
$P_2$ & $^{+0}_{-0}$               & $^{+0}_{-0}$               & $^{+0}_{-0}$               & $^{+0.001}_{-0}$        & $^{+0}_{-0}$ 
\\[1mm]
$P_3$ & $^{+0.015}_{-0.024}$ & $^{+0}_{-0.013}$        & $^{+0.091}_{-0.063}$ & $^{+0.057}_{-0.027}$ & $^{+0.008}_{-0.003}$ 
\\[1mm]
$F_+$ & $^{+0}_{-0.02}$          & $^{+0}_{-0.034}$        & $^{+0}_{-0.054}$          & $^{+0.004}_{-0}$       & $^{+0}_{-0}$ 
\\[1mm]
$F_0$ & $^{+0.003}_{-0.023}$ & $^{+0.007}_{-0.017}$ & $^{+0.036}_{-0.033}$  & $^{+0.028}_{-0.025}$ & $^{+0.008}_{-0}$ 
\\[1mm]
$A_\text{FB,EC}^{z,z'}$
           & $^{+0.05}_{-0.042}$    & $^{+0}_{-0.1}$            & $^{+0.078}_{-0.006}$  & $^{+0.054}_{-0.003}$  & $^{+0.008}_{-0.041}$ 
\\[1mm]
$A_\text{FB}^{x'}$
           & $^{+0}_{-0}$                 & $^{+0}_{-0}$               & $^{+0}_{-0}$                 & $^{+0}_{-0}$               & $^{+0}_{-0}$ 
\\[1mm]
$A_\text{FB}^{y'}$
           & $^{+0}_{-0}$                 & $^{+0}_{-0}$               & $^{+0.002}_{-0}$          & $^{+0}_{-0}$               & $^{+0}_{-0}$      
\\[1mm]
$A_\text{FB}^{x'z'} $
           & $^{+0}_{-0}$                 & $^{+0}_{-0}$               & $^{+0}_{-0}$                 & $^{+0}_{-0}$                & $^{+0}_{-0}$  
\\[1mm]
$ A_\text{FB}^{y'z'}$
           & $^{+0.008}_{-0}$          & $^{+0}_{-0.002}$        & $^{+0.023}_{-0}$          & $^{+0.01}_{-0}$           & $^{+0}_{-0}$  
\\[1mm]
Total    & $^{+0.054}_{-0.057}$   & $^{+0.008}_{-0.11}$   & $^{+0.13}_{-0.09}$       & $^{+0.095}_{-0.038}$  & $^{+0.026}_{-0.041}$ 
\end{tabular}
\caption{Variation in the entanglement measures for different subsystems when the input observables (first column) are allowed to vary within their experimental systematic uncertainty~\cite{ATLAS:2022vym,CMS:2014uod,ATLAS:2017ygi}. Entries smaller than $10^{-3}$ are written as zero. For $A_\text{FB,EC}^{z,z'}$ a measurement is not available, and we assume a systematic uncertainty of $\pm 0.1$.}
\label{tab:syst}
\end{center}
\end{table}

\section{Discussion}

In this Letter we have investigated the bipartite and tripartite entanglement between the spins of the top quark decay products $W$, $b$ and their orbital angular momentum $L$. The entanglement could be determined from experimental data already recorded data at the LHC, and without any model-dependent assumptions. The only missing piece of information is a relative phase $e^{i \alpha}$ between amplitudes with $b$ quark helicities $\pm 1/2$. As shown, this phase has a very small impact on the entanglement and would hardly influence the interpretation of experimental measurements.

The reader may already have noticed two conspicuous features that arise in our theoretical framework. The $LW$ density operator, obtained after trace on the $b$ spin degrees of freedom, is still sensitive to the phase $e^{i \alpha}$.  This is a consequence of the ambiguity in the direct determination of the density operators involving $L$, which we solve by expressing them in terms of $a_{\lambda_1 \lambda_2}$. Even more striking is the fact that the $L$-$b$ entanglement can be established when the $b$ quark spin is not directly measured. This is possible due the change of basis used to write the $LWb$ density operator in terms of a minimal set of parameters $a_{\lambda_1 \lambda_2}$, and the fact that $\lambda = 1/2$ amplitudes are suppressed.

In our estimation of systematic uncertainties for entanglement measures we have used ATLAS and CMS analyses that in some cases use small datasets from LHC Run 1. We remark that with higher statistics, as those available with Run 2 data, the full distributions can be used for measurements, not only asymmetries, and the systematic uncertainties are bound to improve. Aiming at a combined measurement of all observables, a deconvolution of the detector effects for measurements of the four-dimensional distribution has also been proposed~\cite{Boudreau:2013yna,Boudreau:2016pdi}, which also takes into account the correlations among measurements.

In summary, the determination of the bipartite and tripartite entanglement between the spins of top quark decay products and $L$ is based on observables which, except one, have already been measured by the ATLAS or CMS Collaborations, in separate analyses. The current experimental uncertainties suggest that a dedicated analysis of Run 2 data could establish the entanglement between $L$ and the $W$ boson spin, as well as genuine tripartite entanglement, with more than 5 standard deviations. This would be quite a novel type of measurement, which at present has only been achieved with photons. 

\section*{Acknowledgements}
This work of has been supported by the Spanish Research Agency (Agencia Estatal de Investigaci\'on) through projects PID2019-110058GB-C21, PID2022-142545NB-C21 and CEX2020-001007-S funded by MCIN/AEI/10.13039/501100011033, and by Funda\c{c}{\~a}o para a Ci{\^e}ncia e a Tecnologia (FCT, Portugal) through the project CERN/FIS-PAR/0019/2021.

\end{document}